# Degrees of mutual coherence of a 3D polarization state

**José J. Gil**
*Universidad de Zaragoza. Pedro Cerbuna 12, 50009 Zaragoza, Spain*
*ppgil@unizar.es*

**Abstract**

The absolute values of the three degrees of mutual coherence between the analytic signals representing the components of the electric field of a given three-dimensional (3D) polarization state are relative quantities that depend on the laboratory reference frame considered. The extremal values for the said absolute values are determined and analyzed. The reduction to the well-known conventional 2D case is retrieved in a natural way.

## 1. Introduction

In a classic paper [1], Wolf demonstrated that the absolute value $|\mu|$ of the degree of mutual coherence of a 2D polarization state (characterized by its corresponding 2x2 coherency matrix, or polarization matrix) attains its maximum value $|\mu|_{\max} = P$ ($P$ being the degree of polarization) when the laboratory reference system *XY* is transformed to another where the semiaxes of the characteristic polarization ellipse form an angle of 45º with respect to it. This important result can be extended to the study of the extremal absolute values of the degrees of mutual coherence of a general three-dimensional polarization state (i.e. whose direction of propagation is not necessarily stable in time).

## 2. Theoretical background

Next, we summarize a series of concepts that constitute the theoretical background for the formulation and interpretation of the new results.

The polarization matrix (or coherency matrix) **R** of a 3D state of polarization is defined as $\mathbf{R} = \langle \boldsymbol{\varepsilon}(t) \otimes \boldsymbol{\varepsilon}^{\dagger}(t) \rangle$ where $\boldsymbol{\varepsilon}(t)$ is the *3D instantaneous Jones vector* [2] (whose three components are the respective analytic signals of the components of the electric field of the wave), $\otimes$ stands for the Kronecker product, the superscript $^{\dagger}$ denotes conjugate transposed, and the brackets indicate time averaging over the measurement time, so that **R** determines all the second-order polarization properties of an electromagnetic wave at a given point **r** in space.

An overall measure of the polarimetric purity of a 3D state (i.e. of the closeness to a pure state) is given by the 3D *degree of polarimetric purity*, defined as [3-5]

$$P_{3D} = \sqrt{\frac{1}{2}\left(\frac{3\,\mathrm{tr}(\mathbf{R}^2)}{(\mathrm{tr}\,\mathbf{R})^2} - 1\right)}. \tag{1}$$

It is also worth to bring up the *trivial decomposition* (also called the *characteristic decomposition*) [2]

$$\mathbf{R} = P_1 I \hat{\mathbf{R}}_p + (P_2 - P_1) I \hat{\mathbf{R}}_m + (1 - P_2) I \hat{\mathbf{R}}_{u-3D},$$

$$\hat{\mathbf{R}}_p \equiv \mathbf{U} \operatorname{diag}(1,0,0) \mathbf{U}^\dagger, \; \hat{\mathbf{R}}_m \equiv \frac{1}{2} \mathbf{U} \operatorname{diag}(1,1,0) \mathbf{U}^\dagger, \; \hat{\mathbf{R}}_{u-3D} \equiv \frac{1}{3} \mathbf{I}, \quad (2)$$

where, $I \equiv \operatorname{tr} \mathbf{R}$ is the intensity, $\hat{\mathbf{R}} \equiv \mathbf{R}/\operatorname{tr}\mathbf{R}$, $\mathbf{I}$ is the identity matrix and $P_i$ are the so-called *indices of polarimetric purity* (IPP) defined as [6,7]

$$P_1 = \frac{\lambda_1 - \lambda_2}{\operatorname{tr} \mathbf{R}}, \; P_2 = \frac{\lambda_1 + \lambda_2 - 2\lambda_3}{\operatorname{tr} \mathbf{R}}, \quad (3)$$

in terms of the ordered eigenvalues $\lambda_i$ ($\lambda_1 \geq \lambda_2 \geq \lambda_3$) of $\mathbf{R}$; $P_1$ being the *degree of polarization* (i.e., the relative portion of power corresponding to the totally polarized component; note that for 2D states $P_1$ becomes the usual 2D degree of polarization) and $P_2$ being the *degree of two-dimensional polarization* (i.e., the relative portion which is not totally depolarized). The pure state $\hat{\mathbf{R}}_p$ of the characteristic decomposition is called the *characteristic component* of the state $\mathbf{R}$, while its associated well-defined polarization ellipse is called the *characteristic polarization ellipse*. Thus, the IPP determine the structure of polarimetric purity of $\mathbf{R}$ with respect to the relative weights of its incoherent components in the characteristic decomposition. Furthermore, since $P_1$ and $P_2$ are defined from $\lambda_i$, both IPP are invariant under unitary transformations $\mathbf{WRW}^\dagger$ ($\mathbf{W}^\dagger = \mathbf{W}^{-1}$).

Observe that, in accordance with the above considerations and with the arguments of Ellis et al. [8,9], we consider preferable to use the term *degree of polarization* for $P_1$ instead of for $P_{3D}$. A geometric representation of the polarization states in terms of the values of the IPP, as well as the analysis of the regions of the feasible space determined by the axes $P_1, P_2$ can be found in Refs. [10,7].

Despite the invariant information provided by the IPP, it should be stressed that only the unitary transformations that are orthogonal preserve the state of polarization. That is, any proper-orthogonal transformation $\mathbf{QRQ}^T$ ($\mathbf{Q}^T = \mathbf{Q}^{-1}$ with $\det \mathbf{Q} = +1$) can be considered as a representation of the same state as $\mathbf{R}$ but referred to a rotated laboratory reference frame. Thus, concerning the properties of a given polarization state that are invariant with respect to rotations of the reference frame, it is worth to recall that there always exists an orthogonal matrix $\mathbf{Q}_O$ such that the transformed coherency matrix $\mathbf{R}_O$ (called the *intrinsic coherency matrix*) adopts the particular form [11]

$$\mathbf{R}_O \equiv \mathbf{Q}_O \mathbf{R} \mathbf{Q}_O^T = \begin{pmatrix} a_1 & -i\,n_{O3}/2 & i n_{O2}/2 \\ i\,n_{O3}/2 & a_2 & -i n_{O1}/2 \\ -i\,n_{O2}/2 & i n_{O1}/2 & a_3 \end{pmatrix}, \quad (4)$$

in terms of the three *principal intensities* $a_1, a_2, a_3$, with the choice $a_1 \geq a_2 \geq a_3$ (note that $a_i \geq 0$), and the three components $(n_{O1}, n_{O2}, n_{O3})$ of the spin angular momentum vector. The Cartesian reference axes $X_O Y_O Z_O$ corresponding to $\mathbf{R}_O$ constitute the *intrinsic reference frame* [12].

The principal intensities are directly related to so meaningful quantities like [12]: the intensity $I = \operatorname{tr} \mathbf{R} = a_1 + a_2 + a_3$ (i.e., the power density flux of the wave); the degree of linear polarization $P_l \equiv (a_1 - a_2)/I$ (i.e., a measure of how close is the state $\mathbf{R}$ to a linearly polarized one), and the *degree of directionality* $P_d \equiv (a_1 + a_2 - 2a_3)/I$ (i.e., a measure of the lack of randomness in the direction of propagation of the polarization state). Moreover, let us recall that the degree of circular polarization $P_c$ (i.e., a measure of how close is the state $\mathbf{R}$ to a circularly polarized one; $P_c = \pm 1$ for right- and left- circularly polarized states) is given by $P_c = \hat{n} \equiv \sqrt{n_{O1}^2 + n_{O2}^2 + n_{O3}^2}/I$ (note that $\hat{n}$ is invariant under orthogonal transformations, and thus $\hat{n} = \hat{n}_O$).

The degree of polarimetric purity $P_{3D}$ can be expressed as follows in terms of the above-mentioned rotationally invariant quantities $P_l$, $P_c$ and $P_d$ [12]

$$P_{3D} = \sqrt{\frac{3}{4}\left(P_l^2 + P_c^2\right) + \frac{1}{4}P_d^2}, \tag{5}$$

with $P_l^2 + P_c^2 \leq P_d^2$.

## 3. Extremal absolute values for the degrees of mutual coherence of a 3D polarization state

The degrees of mutual coherence

$$\left|\mu_{ij}\right| = \sqrt{\left\langle \varepsilon_i \varepsilon_j^\dagger \right\rangle \left\langle \varepsilon_i^\dagger \varepsilon_j \right\rangle / \left\langle \varepsilon_i \varepsilon_i^\dagger \right\rangle \left\langle \varepsilon_j^\dagger \varepsilon_j \right\rangle} \tag{6}$$

between the analytic signals $\varepsilon_1$, $\varepsilon_2$ and $\varepsilon_3$ of the field components of a 3D state of polarization are relative quantities that depend on the reference frame considered.

Regarding unitary transformations, let us recall that there always exists an unitary transformation $\mathbf{U}\mathbf{R}\mathbf{U}^\dagger = \text{diag}(\lambda_1, \lambda_2, \lambda_3)$ that diagonalizes $\mathbf{R}$, so that $\left|\mu_{ij}\right| = 0$ for the transformed 3D state of polarization represented by the diagonal coherency matrix $\text{diag}(\lambda_1, \lambda_2, \lambda_3)$. It should be noted that unitary transformations preserve the intensity and other invariant quantities derived from the eigenvalues $\lambda_i$ of $\mathbf{R}$ (like, for instance $P_1$, $P_2$ and $P_{3D}$), but they do not preserve, in general, other characteristic properties of the polarization state $\mathbf{R}$ (like, for instance, the shape of the characteristic polarization ellipse).

Moreover, there always exists an orthogonal transformation $\mathbf{R}_E = \mathbf{Q}_E \mathbf{R} \mathbf{Q}_E^T$ such that the transformed coherency matrix $\mathbf{R}_E$ has equal diagonal elements, and as shown below, the maximum achievable values of $\left|\mu_{ij}\right|$ with respect to arbitrary unitary transformations correspond to $\mathbf{R}_E$.

Let us now focus our analysis on orthogonal transformations of $\mathbf{R}$, because they are what preserve the polarization state (i.e., the polarization properties up to a rotation of the laboratory reference frame *XYZ*) and let us consider the limits for $\left|\mu_{ij}\right|$ with respect to arbitrary rotations of *XYZ*.

The minimum values for $\left|\mu_{ij}\right|$ with respect to arbitrary rotations in the real space are achieved when the reference frame is precisely the intrinsic one $X_O Y_O Z_O$, where the differences between the strengths of the components of the electric field are maximum, so that (provided $a_3 > 0$)

$$\left|\mu_{12}\right|_{\min}^2 = \frac{n_{O3}^2}{4a_1 a_2}, \quad \left|\mu_{13}\right|_{\min}^2 = \frac{n_{O2}^2}{4a_1 a_3}, \quad \left|\mu_{23}\right|_{\min}^2 = \frac{n_{O1}^2}{4a_2 a_3}, \tag{7}$$

and therefore the intrinsic coherency matrix $\mathbf{R}_O$ has the property that $\left|\mu_{ij}\right|$ reach their minimum achievable values $\left|\mu_{ij}\right|_{\min}$ with respect to rotations of the reference frame. The limiting case where $a_3 = 0$, i.e. $P_d = 1$, corresponding to a 2D polarization state is considered below in a separate section.

Note that when the state of polarization is pure, $P_1 = P_{3D} = 1$, then $\left|\mu_{ij}\right|_{\min}$ coincide with their maximum and are equal to unity $\left|\mu_{12}\right|_{\min} = \left|\mu_{13}\right|_{\min} = \left|\mu_{23}\right|_{\min} = 1$.

As indicated above, there always exists an orthogonal transformation $\mathbf{R}_E = \mathbf{Q}_E \mathbf{R} \mathbf{Q}_E^T$ such that $\mathbf{R}_E$ has equal diagonal elements; that is, the differences between the diagonal elements of $\mathbf{R}_E$ are zero and thus they reach their maximum values. For a given state $\mathbf{R}$, $\mathbf{Q}_E$ depends on the intrinsic parameters $(a_1, a_2, a_3, n_{O1}, n_{O2}, n_{O3})$ in a complicated manner, but in practice $\mathbf{R}_E$ can be calculated through the Bendel–Mickey algorithm [13] and, provided that $a_3 > 0$, the maximum $|\mu_{ij}|_{\max}$ of the polarization state considered are given by

$$|\mu_{ij}|_{\max} = |r_{E_{ij}}|/\sqrt{|r_{E_{ii}}||r_{E_{jj}}|}, \tag{8}$$

where $r_{E_{ij}}$ are the elements of $\mathbf{R}_E$.

Concerning the measurement of $\mu_{ij}$, let us note that Ellis and Dogariu [14] defined appropriate three-dimensional optical polarimetric arrangements generating nine independent intensity signals that allow the complete determination of $\mathbf{R}$ (hence including $\mu_{ij}$). The calculation (from $\mathbf{R}$) of the extremal values for $|\mu_{ij}|$ is straightforward through the mathematical procedures indicated above.

## 4. Extremal values for the degree of mutual coherence of a 2D polarization state

When the direction of propagation at the point $\mathbf{r}$ considered is stable (i.e., $P_d = 1$ [12]), then $a_3 = 0$ (recall that $\mathbf{R}_O$ has been defined with the choice $a_1 \geq a_2 \geq a_3$), the intrinsic axis $Z_O$ coincides with the direction of propagation, and the intrinsic coherency matrix adopts the particular form

$$\mathbf{R}_O = \begin{pmatrix} a_1 & -in/2 & 0 \\ in/2 & a_2 & 0 \\ 0 & 0 & 0 \end{pmatrix} = \frac{I}{2}\begin{pmatrix} 1+P_l & -iP_c & 0 \\ iP_c & 1-P_l & 0 \\ 0 & 0 & 0 \end{pmatrix}, \tag{9}$$

where the upper-left 2x2 submatrix $\mathbf{\Phi}_O$ represents the 2x2 intrinsic coherency matrix for 2D polarization states. The generic 2x2 coherency matrix $\mathbf{\Phi}$ is obtained through a rotation of an arbitrary angle $\theta$ about the $Z_O$ axis

$$\mathbf{\Phi} = \begin{pmatrix} \cos\theta & -\sin\theta \\ \sin\theta & \cos\theta \end{pmatrix} \mathbf{\Phi}_O \begin{pmatrix} \cos\theta & \sin\theta \\ -\sin\theta & \cos\theta \end{pmatrix}$$
$$= \frac{I}{2}\begin{pmatrix} 1+P\cos 2\varphi \cos 2\chi & P(\sin 2\varphi \cos 2\chi - i\sin 2\chi) \\ P(\sin 2\varphi \cos 2\chi + i\sin 2\chi) & 1-P\cos 2\varphi \cos 2\chi \end{pmatrix}, \tag{10}$$

where $P \equiv P_l$ is the degree of polarization, while $\varphi = \theta/2$ and $\chi$ are precisely the respective azimuth and ellipticity angle of the characteristic polarization ellipse of the 2D state. Moreover, the elements $\phi_{ij}$ $(i.j = 1, 2)$ of $\mathbf{\Phi}$ can be written as follows in terms of the corresponding standard deviations $\sigma_1, \sigma_2$ and the complex degree of mutual coherence $\mu$

$$\mathbf{\Phi} = \begin{pmatrix} \sigma_1^2 & \mu\sigma_1\sigma_2 \\ \mu^*\sigma_1\sigma_2 & \sigma_2^2 \end{pmatrix}, \tag{11}$$

where

$$\sigma_1^2 \equiv \phi_{11} = \langle |\varepsilon_1(t)|^2 \rangle, \qquad \sigma_2^2 \equiv \phi_{22} = \langle |\varepsilon_2(t)|^2 \rangle, \qquad \mu = \frac{\phi_{12}}{\sigma_1 \sigma_2} = \frac{\phi_{12}}{\sqrt{\phi_{11}\phi_{22}}}. \tag{12}$$

Let us now observe that $|\mu|$ can be expressed as follows in terms of the degree of polarization $P$, the azimuth $\varphi$ and the ellipticity angle $\chi$ of the characteristic component of $\mathbf{\Phi}$

$$|\mu| = \sqrt{\frac{\phi_{12}\,\phi_{21}}{\phi_{11}\,\phi_{22}}} = \sqrt{\frac{P^2 - P^2 \cos^2 2\varphi \cos^2 2\chi}{1 - P^2 \cos^2 2\varphi \cos^2 2\chi}} = \sqrt{\frac{P_c^2 + P_l^2 \sin^2 2\varphi}{1 - P_l^2 \cos^2 2\varphi}} \qquad (13)$$

so that, as shown by Wolf [1], $0 \le |\mu| \le P$. In accordance with the generic results for 3D states, $|\mu|$ vanishes for 2D unpolarized states $(P = 0 \Rightarrow |\mu| = 0)$, while for 2D pure states the equality $|\mu| = P = 1$ holds necessarily. For mixed states, the maximum $|\mu|_{max} = P$ corresponds either to states whose characteristic component is circularly polarized $(|P_c| = 1)$, or to states whose characteristic polarization ellipse has azimuth $\varphi_1 = \pi/4$ or $\varphi_2 = 3\pi/4$; that is, $|\mu| = P$ for the Cartesian reference frames $X_{E1}Y_{E1}$ or $X_{E2}Y_{E2}$ in which the strengths of the two orthogonal components are equal. Consequently, the 2x2 coherency matrix adopts the following forms when referred to those particular reference frames

$$\boldsymbol{\Phi}_{E_1} = \frac{1}{2}\begin{pmatrix} a_1 + a_2 & a_1 - a_2 - in \\ a_1 - a_2 + in & a_1 + a_2 \end{pmatrix} = \frac{I}{2}\begin{pmatrix} 1 & P_l - iP_c \\ P_l + iP_c & 1 \end{pmatrix}$$

$$\boldsymbol{\Phi}_{E_2} = \frac{1}{2}\begin{pmatrix} a_1 + a_2 & -a_1 + a_2 - in \\ -a_1 + a_2 + in & a_1 + a_2 \end{pmatrix} = \frac{I}{2}\begin{pmatrix} 1 & -P_l - iP_c \\ -P_l + iP_c & 1 \end{pmatrix} \qquad (14)$$

Moreover, from Eq. (13), the minimum value of $|\mu|$ for a given polarization state is reached when the polarization ellipse has azimuth $\varphi = 0, \pi/2$ so that $|\mu|_{min}$ is given by

$$|\mu|_{min} = \sqrt{\frac{P^2 \sin^2 2\chi}{1 - P^2 \cos^2 2\chi}} = \sqrt{\frac{P_c^2}{1 - P_l^2}} \qquad (15)$$

Thus $|\mu| = |\mu|_{min}$ when $\boldsymbol{\Phi}$ is referred to the intrinsic reference frame $X_O Y_O$ or to a reference frame orthogonal to it.

Note that, unlike $|\mu|_{max}$, $|\mu|_{min}$ depends on the ellipticity angle $\chi$ and takes values in the range $0 \le |\mu|_{min} \le P$. The lower limit $|\mu|_{min} = 0$ corresponds to states with zero angular momentum (which include both unpolarized states and mixed states whose characteristic component has linear polarization). The upper limit $|\mu|_{min} = P = |\mu|_{max}$ corresponds to mixed states whose characteristic component has circular polarization (i.e., $\chi = \pm \pi/4$). For the intermediate case of a mixed state, whose characteristic component has an elliptic polarization with $\chi = \pi/8$, then $|\mu|_{min} = P/\sqrt{2 - P^2}$.

Obviously, $|\mu|$ as well as their extremal values can easily be determined through conventional Stokes polarimetry.

### Acknowledgement

The author thanks Dr. Alfredo Luis for insightful discussions on this subject. This research was supported by Ministerio de Economía y Competitividad and European Union, grants FIS2011-22496 and FIS2014-58303-P.